\documentstyle [english,epsf,12pt]{article}
\begin{document}
 
\newcommand{\BS}{\bigskip}
\newcommand{\SECTION}[1]{\BS\begin{center}{\large\section{\bf #1}}\end{center}}
\newcommand{\SUBSECTION}[1]{\BS{\large\subsection{\bf #1}}}
\newcommand{\NOTE}[1]{\marginpar{\small #1}}
\newcommand{\ABS}[1]{\mid\!#1\!\mid}
\newcommand{\IVPB}{\rm pb^{-1}}
\newcommand{\IVNB}{$\rm nb^{-1}$}
\newcommand{\EE}{e^+e^-}
\newcommand{\PGG}{\pi^0 \rightarrow \gamma \gamma}
\newcommand{\MEN}{\mu^+ \rightarrow e^+ \nu \bar{\nu}}
\newcommand{\KPKM}{K^+ K^-}
\newcommand{\KK}{K^0-\bar{K}^0}
\newcommand{\KM}{K^+ \rightarrow \mu^+ \nu_\mu}
\newcommand{\KPM}{K^+\rightarrow\pi^0\mu^+\nu_\mu}
\newcommand{\KMG}{K^+ \rightarrow \mu^+ \nu_\mu \gamma}
\newcommand{\KMM}{K^- \rightarrow \pi^0 \mu^- \bar{\nu_\mu}}
\newcommand{\KPO}{K^+ \rightarrow \pi^+ \pi^0}
\newcommand{\KPPP}{K^+ \rightarrow \pi^+ \pi^0 \pi^0}
\newcommand{\KPTRE}{K^+_{\mu 3}}
\newcommand{\KPDUE}{K^+_{\mu 2}}
\newcommand{\KOPM}{K^0_L \rightarrow \pi^- \mu^+ \nu_\mu}
\newcommand{\FP}{f_+(q^2)}
\newcommand{\FM}{f_-(q^2)}
 
\hyphenation{multi-pli-ci-ties}
\textheight 650pt \textwidth 400pt
\setlength{\unitlength}{0.01pt}
\thispagestyle{empty}
\vspace*{0.4cm}
\vspace*{0.2cm}
\rightline{\large ROM2F/96/13}
\vspace*{0.2cm}
\rightline{5 March 1996}
\vspace*{0.4cm}
\begin{center}
\Huge {\bf
Measurement of muon polarization \\
in $\KPM$ at a $\phi$ factory}
\vspace*{0.4cm}
\vskip 2cm
\large {
                   Paolo Privitera\\
{\small Dipartimento di Fisica, Universit\`a degli Studi di Roma II "Tor Vergata"} \\
   {\small  and INFN, Sezione di Roma II, Italy}
}
\vskip 3cm
 
\end{center}
\begin{abstract}
The potentiality of an experiment measuring the muon polarization in 
\mbox{$\KPM$} 
at a $\phi$ factory is discussed, with particular attention to the 
possibility of revealing unexpected violation of Time reversal invariance
through the transverse polarization component.
An experimental method is proposed which is based on the reconstruction
of the $\KPM$ decay products after stopping 
the charged kaons in a target placed around the beam pipe.
The muon polarization is measured by observing the asymmetry in the angular
 distribution of positrons from the $\MEN$ decay of muons stopped in 
a polarization analyser. 
It is shown that an error $\sigma_P\simeq 5\cdot 10^{-4}$ on each muon
polarization component can be obtained,
 which represents a factor ten improvement
with respect to current limits  on T invariance violation in the $\KPM$ decay
and a precise measurement of the $Re\xi$ parameter at the level of 
$2\cdot 10^{-3}$.
\end{abstract}
\vskip 2cm
\pagebreak
\begin{titlepage}
\mbox{}
\end{titlepage}
\pagebreak

\setcounter{page}{1}
\SECTION{\label{intro} Introduction}
A precision study of the CP violation parameters in the $\KK$ 
system is the central item of the physics program at a $\phi$ factory 
\cite{Handb}.
On the other hand, a $\phi$ factory will produce a large statistics 
($\simeq 10^{10}$ per year) of $\KPKM$ pairs. 
Significant improvements in the understanding of the decay and interaction 
of charged kaons are expected from the approved experiments \cite{KLOE}
\cite{FINUDA} at the Frascati $\phi$ factory \cite{DAFNE}.   

In this context, 
a measurement of muon polarization in the semileptonic decay $\KPM$ ($\KPTRE$)
would be very interesting.  
In fact, the component of the muon polarization transverse to the decay plane
 is sensitive to possible violation of Time reversal (T) invariance, as first 
noticed by Sakurai \cite{Sakurai}.   
The transverse muon polarization
\begin{eqnarray}
      P_T = \frac{ \vec{s}_{\mu^+} \cdot ( \vec{p}_{\pi^0} \times
   \vec{p}_{\mu^+} ) }{| \vec{p}_{\pi^0} \times \vec{p}_{\mu^+} | },
\label{eq:pt}
\end{eqnarray}
in which $\vec{s}_{\mu^+}$ is the muon spin vector, $\vec{p}_{\mu^+}$ and 
$\vec{p}_{\pi^0}$ are the muon and neutral pion momenta, changes sign 
under the T-reversal operation. A non-zero value of $P_T$ would 
evidentiate a T violation. 

In the standard model, where CP violation is introduced
with an imaginary phase in the Cabibbo-Kobayashi-Maskawa matrix elements 
\cite{Ckm}, $P_T$ is expected to be very small ($\simeq 10^{-6}$). 
 Also, the presence of electromagnetic interaction (even if T 
invariant) in the final state can produce a non-zero $P_T$,
 estimated to be $10^{-6}$ for $\KPTRE$ \cite{Fin}.
Thus, the measurement of a non-zero value ($>10^{-6}$) of $P_T$ would be a 
definite sign of physics beyond the standard model. 
On the other hand, there are several models \cite{mod1} \cite{mod2} containing
charged Higgs particles or other non-standard scalar interactions like 
lepto-quarks that do not violate existing experimental constraints from 
the neutron dipole moment and $\epsilon'/\epsilon$ and allow a transverse
muon polarization as large as $P_T \simeq 10^{-3}$.

The magnitude of the other two components of the muon polarization is
 determined by the form factors describing the decay $\KPM$. 
A precise measurement of the total muon polarization would allow the
verification of predictions of the modern chiral symmetry theory \cite{Chir}.

In this paper, the feasibility of an Experiment for T Invariance in KAon decay
(ETIKA) at a $\phi$ factory is analysed. Section \ref{theor} is devoted to the
description of the 
theoretical framework of the $\KPM$ decay. 
The achievable accuracy of the muon polarization measurement at a $\phi$ factory
 is analysed in Section \ref{stat}.
In Section \ref{phifact} the 
essential properties and requirements of the ETIKA experiment are discussed. 
Other physics topics are presented in Section \ref{phtpo}.
Lastly, the conclusions are given.  

\begin{figure}[p]
\vspace{-4cm}
\hspace*{1.5cm}
\mbox{\epsfysize20.0cm\epsffile{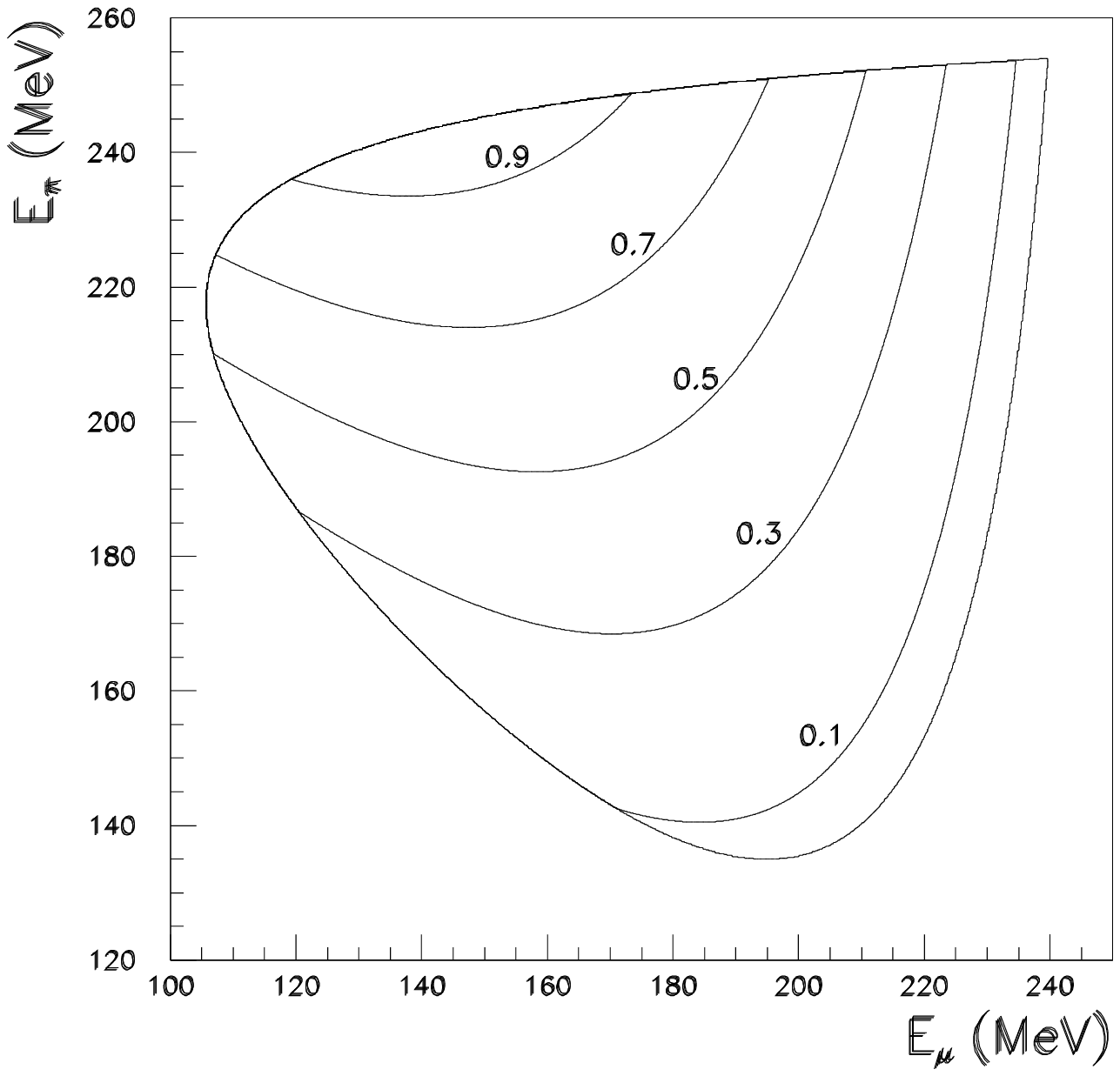}}
\vspace*{-4cm}
\caption[Dalitz plot K]{Contours indicating the relative phase space population
of the $\KPTRE$ Dalitz plots}
\label{fig:Dalitz}
\end{figure}

\SECTION{\label{theor} Theoretical Framework}
In the usual V-A theory of the $\KPTRE$ decay, the matrix element governing the 
decay is proportional to $<\pi | J_\lambda | K >$, in which $J_\lambda$ is the 
strangeness-changing hadronic current. The hadronic vertex associated to the
 current is a function of three four-vectors, only two of which are independent 
by energy-momentum conservation. The combinations $P_K-P_\pi$ and $P_K+P_\pi$
are usually chosen as basis vectors, where $P_K$ and $P_\pi$ are the 
kaon and pion four-vectors, respectively. 
Also, there is only one independent 
scalar, apart from the kaon and pion masses, which can be formed from the 
basis vectors. It is usually chosen to be $q^2=(P_K-P_\pi)^2$. 
Being $<\pi | J_\lambda | K >$ a four-vector which represents the hadronic vertex, 
it must be expressible as a linear combination of the basis vectors, with 
coefficients at most depending on $q^2$. 
The standard expression is 
\begin{equation}
<\pi | J_\lambda | K > = \FP (P_K+P_\pi)_\lambda + \FM (P_K-P_\pi)_\lambda.  
\end{equation}
In this formalism, all experimental quantities except absolute rates can be 
written as a function of only one complex parameter $\xi (q^2)$ given by the 
ratio of $\FM$ to $\FP$. 
The Dalitz plot density in the kaon rest frame\footnote{For the sake of 
simplicity, the notation $K,~\mu $ and $\pi$ instead of $K^+,~\mu^+$ and 
$\pi^0$ will be used in the following.}
 is then given by \cite{Daltz} 
\begin{eqnarray}
\frac{d^2 N}{d E_\pi d E_\mu} \propto |\FP|  ( A + B Re\xi + C |\xi|^2),
\label{for:Dalitz}
\end{eqnarray}
in which 
\begin{eqnarray}
A &=& m_K (2 E_\mu E_\nu - m_K E_{\pi}')+m_\mu^2 (\frac{1}{4}E_\pi' -E_\nu),\nonumber 
\\
B &=& m_\mu^2 (E_\nu-\frac{1}{2}E_\pi'), \nonumber \\
C &=& \frac{1}{4} m_\mu^2 E_\pi', \nonumber \\
E_\pi' &=& E_\pi^{max}-E_\pi,~~~~~
 E_\pi^{max}= \frac{m_K^2+m_\pi^2-m_\mu^2}{2m_K}
, \nonumber 
\end{eqnarray}
in which $m_i$ and $E_i$ are the energy and mass of particle $i$.  
The Dalitz plot density of the $\KPM$ decay, corresponding to equation 
(\ref{for:Dalitz}) and the Particle Data Group (PDG) \cite{PDG} values 
$Re\xi = -0.35 \pm 0.15$ and $Im \xi = -0.017 \pm 0.025$, is plotted in Fig.
\ref{fig:Dalitz}.

Expressions for the muon polarization in $\KPTRE$ decays have been given by 
several authors \cite{POLMU}.
Since the neutrino from the decay has unit elicity, the muon moving with  a 
given angle and momentum relative to the neutrino has a unique spin orientation.
In the K rest frame, the muon polarization is then a unit vector whose 
components are uniquely specified by the decay kinematics. The expression 
given by Cabibbo and Maksymowics \cite{POLMU} for the muon polarization in 
the kaon rest frame is $\vec{P}=\vec{A}/|\vec{A}|$, in which  
\begin{eqnarray}
\vec{A} = a_1(\xi)\vec{p_\mu}-a_2(\xi)\left[ \frac{\vec{p_\mu}}{m_\mu}\left[(m_K
-E_\pi)+ \frac{\vec{p_\pi}\cdot\vec{p_\mu}(E_\mu-m_\mu)}{|\vec{p_\mu}|^2}\right]+
\vec{p_\pi} \right]+m_K Im\xi (\vec{p_\pi}\times\vec{p_\mu}),
\label{eq:pol}
\end{eqnarray}
with
\begin{eqnarray}
a_1(\xi) &=& 2\frac{m_K^2}{m_\mu}[E_\nu+Reb(q^2)E_\pi'] \nonumber \\
a_2(\xi) &=& m_K^2+2Reb(q^2)m_kE_\mu+|b(q^2)|^2 m_\mu^2, \nonumber \\
b(q^2) &=& \frac{\xi(q^2)-1}{2}, \nonumber
\end{eqnarray}
in which $\vec{p_\mu}$ and $\vec{p_\pi}$ are the momenta of the muon and pion, 
respectively, in the kaon rest frame.

Notice that $\vec{P} = \vec{P}(P_\mu,P_\pi,\xi)$. Having measured the
 kinematics of an event ({\it i.e.} the four-vectors $P_\mu$ and $P_\pi$), 
the actual value of $\vec{P}$ is determined by $\xi$. In particular,
for each event a reference frame can be defined having axes 
$\hat{z}= \vec{p_\mu}/|\vec{p_\mu}|$, $\hat{x}=\vec{p_\pi}\times\vec{p_\mu}/
|\vec{p_\pi}\times\vec{p_\mu}|$, and $\hat{y} = \hat{z}\times\hat{x}$. 
The longitudinal component of the muon polarization in this reference frame is
then $P_L= \vec{P} \cdot\hat{z}$, the perpendicular component is 
$P_P= \vec{P} \cdot\hat{y}$ and the transverse component is 
$P_T= \vec{P} \cdot\hat{x}$.
The $P_T$ component is exactly the T-odd observable identified in equation 
(\ref{eq:pt}). 
According to equation (\ref{eq:pol}), $P_T$ is proportional to $Im \xi$, and  
 if T invariance is conserved one expects $P_T$ to be zero. On the other hand,
the $P_L$ and $P_T$ components are sensitive to $Re\xi$.   
The values averaged over the Dalitz plot are  $<P_L> \simeq 0.75$,
 $<P_P> \simeq -0.50$ and $<P_T> \simeq 0.22 Im \xi$. 

\begin{figure}[p]
\vspace{-4cm}
\hspace*{1.5cm}
\mbox{\epsfysize20.0cm\epsffile{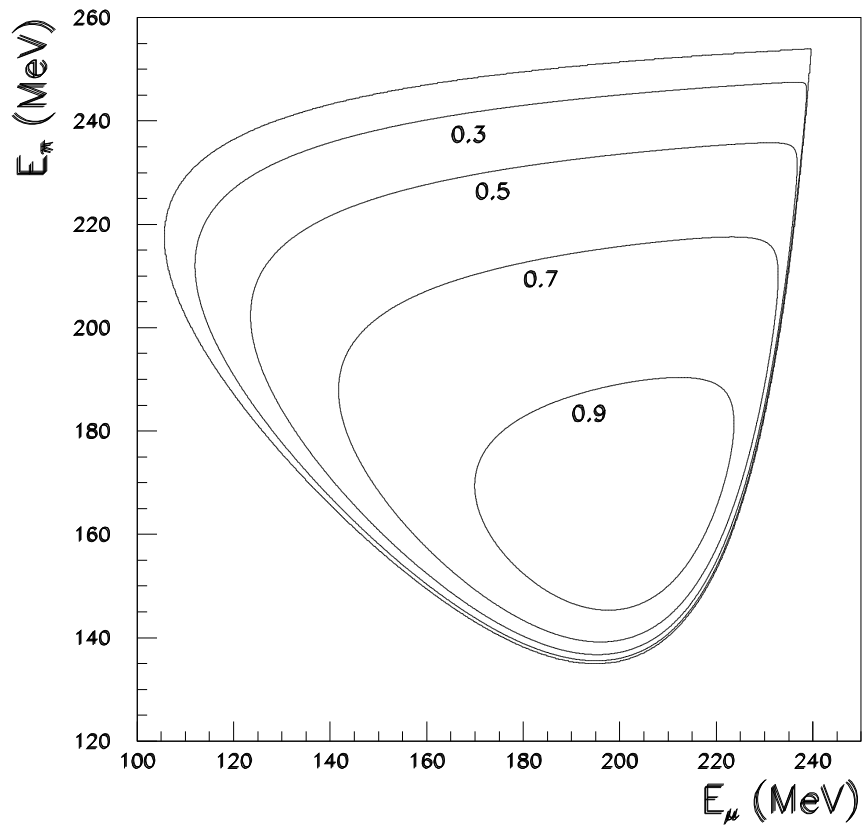}}
\vspace{-4.0cm}
\caption[Polt plot K]{Contours indicating the sensitivity to $\xi$ over the 
 Dalitz plot from muon polarization measurement}
\label{fig:senspolt}
\end{figure}

From the experimental point of view, it is important to identify the regions of
the Dalitz plot where the best sensitivity to $\xi$ is obtained from the 
measurement of the muon polarization. 
For example, one can use the variation of the polarization $|\Delta \vec{P}|$
in a given point
of the Dalitz plot ($E_\mu,E_\pi$) for a fixed change of $Re\xi$ or $Im \xi$.
Since the error on the measurement of a polarization scales as 
$\simeq 1/ \sqrt{N}$, where $N$ is the number of events  in the point 
($E_\mu,E_\pi$) of the Dalitz plot, the ratio 
$|\Delta \vec{P}|/(1/ \sqrt{N})=|\Delta \vec{P}| \sqrt{N}$ is a good estimator
 of the sensitivity. A measure of the sensitivity to $Im \xi$ is
given by $|\Delta P_T| \sqrt{N}$. A measure of the sensitivity to $Re \xi$ is 
given by $|\Delta \vec{P}_{plane}| \sqrt{N}$, where $\vec{P}_{plane}= P_L\hat{z}
+ P_P\hat{y}$ is the component of the muon polarization lying on the decay
 plane. The sensitivity over the Dalitz plot is almost identical for the two 
cases, and contour lines corresponding to this estimator are shown 
in Fig. \ref{fig:senspolt}. 
It is clear that the best sensitivity to both $Im\xi$
 and $Re \xi$ is obtained in the region of low
$\pi^0$ energy and high $\mu^+$ energy. However, the sensitivity to $\xi$
is still significant over most of the Dalitz plot area, allowing a 
non-critical definition of the experimental acceptance.

\SECTION{\label{stat} The Achievable Accuracy}
The DA$\Phi$NE $\phi$ factory is expected to collide $\EE$ beams with a 
luminosity of \mbox{$5\cdot10^{32}$cm$^{-2}$sec$^{-1}$}, which corresponds to 
$\simeq 2.2 \cdot 10^{10}~\phi$ produced in a year of $10^7$ seconds. Thus, 
$\simeq 1.1 \cdot 10^{10}~K^+K^-$ from the $\phi$ decay will be produced in one year
 of running. Assuming a branching ratio of $3.18\%$ \cite{PDG} for $\KPTRE$, 
a number $N=3.5\cdot 10^8$ of $\KPM$ will be available. In order to measure the muon
 polarization, the subsequent spin analysing 
decay $\mu^+ \rightarrow e^+ \nu \bar{\nu}$ must be used. After integration over 
the positron energy spectrum, the decay angular distribution in the muon rest 
frame is:     
\begin{eqnarray}
f(x,P) = \frac{1}{2} \left( 1+ \frac{P}{3}x \right), 
\label{eq:mudec}
\end{eqnarray}
in which $P$ is the muon polarization, and $x=\hat{P}\cdot\hat{p_{e^+}}$ is the 
cosinus of the angle between the polarization unit vector $\hat{P}$ and the unit 
vector along the positron 
momentum. An estimate of the error on the muon polarization obtained using the 
polarization sensitivity of equation (\ref{eq:mudec}) is given by \cite{SIGMA}:
\begin{eqnarray}
\sigma_P  = \frac{1}{\sqrt{N}} \left[ \int_{-1}^{+1}
\frac{1}{f} \left( \frac{\partial f}{\partial P} \right)^2 dx \right]^{-\frac{1}{2}} = 
\frac{1}{S\sqrt{N}}. 
\label{eq:sigma}
\end{eqnarray}
For example, in the case of the T violating polarization $P_T$, which is 
$\simeq 0$, $S\simeq 0.19$ and the expected error is: 
\begin{eqnarray}
\sigma_P  \simeq \frac{1}{0.19 \sqrt{3.5\cdot 10^8}} \simeq 2.8 \cdot 10^{-4}.
\label{eq:sigma1}
\end{eqnarray}
Similar errors are also expected for the $P_L$ and $P_P$ components.

It should be noticed that an improved sensitivity is obtained if the full 
$f(E_{e^+},x,P)$ distribution is used, once appropriate cuts on the positron 
energy $E_{e^+}$ and angle are applied. In fact, it can be shown \cite{IMPR}
 that $S$ can be as large as 0.26, thus reducing the error quoted in equation
 (\ref{eq:sigma1}) to $2.0 \cdot 10^{-4}$.

The best existing measurement of $P_T$ \cite{BEST} is consistent with zero with
 an error of $5 \cdot 10^{-3}$. Measurement of $P_L$ and $P_P$ are quite old 
\cite{PDG} and have large errors. An error $\sigma_P \simeq 5 \cdot 10^{-4}$ would
represent a factor ten improvement on the sensitivity to T violating effect 
in $\KPTRE$. Also, it corresponds to a precise measurement of $Re\xi$, with an error 
$\simeq 2 \cdot 10^{-3}$, through the $P_L$ and $P_P$ components. 
In order to reach such precision, the overall acceptance of the ETIKA 
experiment at the $\phi$ factory should be in the range of $\epsilon = 15-30\%$
 (depending on the actual value of $S$), which includes the fraction of 
accepted $K^+$ from the $\phi$ decay, the efficiency of reconstructing the 
$\KPM$ decay, and the acceptance over the spin analysing decay 
$\mu^+ \rightarrow e^+ \nu \bar{\nu}$. 

If a luminosity of $10^{33}$cm$^{-2}$sec$^{-1}$ should be reached, the quoted errors
will improve by a factor $\sqrt{2}$.
Alternatively, the overall acceptance needed to reach
a precision of $\simeq 5 \cdot 10^{-4}$ would reduce to $8-15\%$.

\SECTION{\label{phifact} A $\phi$ Factory ETIKA Experiment}

The measurement of the muon polarization is performed with the 
standard technique of stopping the positive muons in a non-depolarizing 
material, like Carbon or Aluminium, and then observing the asymmetry in the 
angular distribution of the positrons from  the $\mu^+ \rightarrow e^+ \nu 
\bar{\nu}$ decay. It should be noted that the decay $\phi \rightarrow K^+ K^-$
allows in principle also the study of $\KMM$, when the $K^-$ decay is observed
 in flight. However, stopped negative muons 
are rapidly captured in atomic orbits and almost completely depolarized. Thus, 
the $\KMM$ decay cannot be used for the measurement of the muon polarization 
in charged kaon semileptonic decays, and the ETIKA experiment must be able to   
distinguish between $K^+$ and $K^-$ decay products.
                                                  
A possible solution is the kaon charge identification by tracking particles
 curvature in a magnetic field.
On the other hand, the presence of a magnetic field will cause the spin of 
stopped muons to precess. The net effect will be a significant reduction of the 
sensitivity, with a corresponding increase by a factor of three
of the expected error $\sigma_P$ given by equation (6).
 Moreover, the presence of radial components of the 
field which are typically at the level of a few gauss can introduce systematic
effects which are difficult to control. 
The polarization analyser could be placed in a magnetic shield, but, due to 
the high value of the magnetic field (0.5-1 Tesla), the level of 0.1 gauss
inside the shield needed to reduce the systematic uncertainties would be hardly 
reachable \cite{Ardu}. In conclusion, this approach looks very difficult
and, even if feasible, complicated and costly.

A different experimental method is proposed, 
based on stopping charged kaons in a target surrounding the 
$e^+e^-$ interaction point, which provides
a good acceptance with detectors of reasonable dimensions and the separation  
of $K^+$ and $K^-$ without a magnetic field.
Charged kaon pairs produced at the DA$\Phi$NE $\phi$ factory will have 
 $\simeq$16 MeV kinetic energy, and stop in about 0.5 g/cm$^2$
thickness of material. As a matter of fact, a Carbon target 
 surrounding the DA$\Phi$NE beam pipe is the key element for the study 
of hypernuclear states proposed by the FINUDA \cite{FINUDA} experiment. 
A schematic picture of the ETIKA experiment is shown in 
Fig. 3.
 The Beryllium beam pipe is surrounded by
$\phi$ segmented scintillators which represent the active target.
A positively charged kaon stops into the scintillator target, and its subsequent decays
can then be observed.
 On the other hand, the $K^-$ 
brought to rest into the target suffers nuclear capture, giving rise to
characteristic "stars" from nuclear evaporation.      
Low energy heavy fragments, $\Lambda$ and $\Sigma$ particles emitted in the
 process of $K^-$ absorption 
 give distinct signals in scintillators and tracking detectors placed around 
the target, when compared with the single muon track coming from the 
$K^+$ decay. Also, the prompt signal ($\le \Lambda$ lifetime)
from the products of $K^-$ absorption can be used to efficiently tag the $K^+$
decay having a 12 ns lifetime. 
The energy deposited in the target scintillators by the charged kaons and the back to
back topology of the $\KPKM$ events 
is used for the first level trigger of the experiment.
Also, a tracking detector is 
placed between the scintillator target and the beam 
pipe, in order to precisely determine the $K^+$ stopping
 point into the target. 

Low mass tracking chambers are placed between the target scintillators 
 and the polarization analyser in order to measure precisely the $\mu^+$ 
direction of flight. The polarization analyser detector, in its simplest
configuration, is composed by layers of 
non-depolarysing absorber, interleaved by tracking detector planes, which 
track the muon and determine its range and also measure the positron 
from the $\MEN$ decay.
 A resolution of $\simeq 3$ MeV on the muon energy is obtained 
on the basis of the stopping layer by using thin absorbers 
(for example 5 mm thick Carbon layers).
Photons from the $\pi^0$ decay are reconstructed in the
 electromagnetic calorimeter (not shown in Fig. 3)   
 placed around the polarization analyser detector.  
The ETIKA detector would fit in a cylinder of $\simeq 3$ m diameter and 
$\simeq 3$ m length.

A simulation \cite{Geant}, in which the polarization analyser is formed by
70 Carbon layers with 5 mm thickness,
showed that an overall 
acceptance $\epsilon \simeq 20 \%$ can be obtained. It is 
clear that this value for 
the efficiency $\epsilon$ is only a reasonable estimate, and a detailed Monte
 Carlo simulation of the individual detectors is needed in order to be more
precise. Nevertherless, this estimate of the acceptance is well within the
range quoted in Section~\ref{stat} ($\epsilon \simeq 15-30 \%$), and it 
shows that 
an error $\sigma_P \simeq 5 \cdot 10^{-4}$ on the T violating muon polarization
is achievable with the experimental method proposed.

The effect of detectors resolutions was estimated by 
a Monte Carlo simulation.
Energy and angular resolutions were chosen on the basis of the performances
of available detector techniques. In particular, a resolution of 10 mrad was assumed for both
the polar ($\theta$) and azimuthal ($\phi$) angle reconstruction of positive
 muons, together with 3 MeV resolution in energy. 
 $\sigma_{E_{\gamma}}/E_{\gamma}= 5\%/\sqrt{E_\gamma}$
 and a granularity 
corresponding to 20 mrad angular resolution in $\theta$ and $\phi$ 
was taken for the electromagnetic calorimeter. 
The direction of photons interacting in the
 polarization analyser was assumed to be reconstructed with 70 mrad precision
on both angles. 
The resolution on the positron direction was taken to be 
120 mrad in $\theta$ and $\phi$.
The simulation showed that detector resolutions and the procedure of reconstruction
 of the $\KPM$ kinematics do not spoil the sensitivity to the polarization 
measurement and do not introduce significant systematic effects.

A large part of the systematic uncertainties on the muon polarization 
measurement can be estimated from the data themselves.
For example, the decay $K^+\rightarrow\mu^+ \nu$ ($\KPDUE$) provides a 
copious source of monoenergetic muons,  which will be univoquely
 identified by their long range into the polarization analyser.
 The characteristics of $K^-$ interaction into the target
will be precisely determined looking at the detector signals in events 
tagged by a \mbox{$K^+\rightarrow\mu^+ \nu$} decay.
Also, these events allow a calibration of the muon polarization measurement, 
 since muons from the $\KPDUE$ decay have $P_L=-1$ and zero transverse 
components, and any systematic effect will show up in the polarization sensitive
 positron angular distribution. 
The identification and kinematics recontruction of the $\pi^0$ from the 
$\KPM$ decay can be studied with the two-body decay $\KPO$, which
features a monoenergetic $\pi^+$ and two photons reconstructed 
in the electromagnetic calorimeter or polarization analyser. 
The energies of the photons can be predicted by applying momentum conservation, 
 the photons directions being measured by the detectors. This will allow
a calibration of the detectors energy response to photons. Also,  
the procedure of kinematics reconstruction of events in which one photon 
is measured in the electromagnetic calorimeter and the other photon is measured
 in the polarization analyser will be tested. 

Background events in the sample of $\KPM$ candidates come mainly from other
$K^+$ decays.
 In fact, a pion stopping in the range detector will
decay into a muon of 4 MeV kinetic energy, which will stop typically
 in the same layer where the originating pion stopped, and then decay through
$\MEN$.  
Thus, $\KPO$ or \mbox{$\KPPP$} (in which two of the photons coming from the decays of the 
neutral pions are not reconstructed) will mimic a $\KPM$ event.   
However, these events can be in large part rejected by kinematics constraints 
and requirements on the momentum of the observed particles.
In any case, expected backgrounds will at most reduce the sensitivity to the
muon polarization measurement, but they do not produce any significant
 spurious component of the polarization.

The presence of a magnetic field can introduce systematic effects in the 
polarization measurement. 
In fact, the muon polarization will precess around the field, 
generating spurious components of the polarization.
The expected effects in the case of the ETIKA experiment are
small, since, given the presence of a magnetic field along a fixed
 direction, the muon polarization of muons from the $\KPM$ decay
 will be isotropically distributed around 
this direction, and any spurious component will tend to cancel. 
This statement was checked with a Monte Carlo simulation, which showed 
that enclosing the ETIKA detector in a magnetic shield in order to reduce the 
earth magnetic field (typically a few tenths of gauss) and any other 
spurious field to less than 0.1 gauss, will bring the systematic uncertainty to a
negligible level.

\SECTION{\label{phtpo} Other Physics Topics}

The specificity of the ETIKA experiment can be used 
to measure the muon polarization in the $\KMG$ decay.
Also in this case the transverse component of the muon polarization, proportional 
to $(\vec{p_\mu} \times \vec{p_{\gamma}})\cdot \vec{s_\mu}$, can reveal a possible
violation of the T invariance. The transverse muon polarization in $\KMG$ 
is negligible within the standard model, but the electromagnetic final state 
interaction can mimic T violation effects as large as $10^{-3}$ \cite{Kmg}.
However, the contributions from final state interaction can in principle be 
precisely calculated, thus leaving the possibility of searching for T violation
below the $10^{-3}$ level. Also, the knowledge of the expected 
value of the transverse polarization in the $\KMG$ decay can be used to 
calibrate the polarization measurement. The expected sensitivity of the ETIKA
experiment for the $\KMG$ decay mode is similar to that of the $\KPM$ mode.

Apart from the specific capability of measuring the muon polarization, the 
detector is able to recontruct the kinematics of the kaon decay. 
For example, the Dalitz plot density of $\KPM$ given in equation (3)
 can be studied,
which will provide an independent measurement of $Re\xi$.  
Precise measurements of the branching ratios and Dalitz plot densities of the
kaon decay modes will be performed, as well as a determination of the 
charged kaon lifetime. 

\SECTION{\label{concl} Conclusions}
The potentiality of an experiment measuring the muon polarization in $\KPM$ 
at a $\phi$ factory was discussed. Particular attention was paid to the 
possibility of revealing unexpected violation of Time reversal invariance
through the transverse polarization component.
An experimental method is proposed which is based on the 
 reconstruction the $\KPM$ decay products after stopping  
the charged kaons in a target placed around the beam pipe.
The muon polarization is measured by observing the asymmetry in the angular
 distribution of positrons from the $\MEN$ decay of muons stopped in 
in a polarization analyser. The experiment does not use a magnetic field,
which limits the cost and dimensions of the detector.
It was shown that an error $\sigma_P\simeq 5\cdot 10^{-4}$ on each muon
polarization component can be obtained, which represents a factor ten improvement
with respect to current limits  on T invariance violation in the $\KPM$ decay
and a precise measurement of the $Re\xi$ parameter at the level of 
$2\cdot 10^{-3}$. The systematic uncertainties are expected to be 
negligible. 
The detector is well suited to perform also other precise measurements,
like the muon polarization in $\KMG$ and the study of the charged kaon
decay modes.
The ETIKA experiment could be a very interesting option for the long term 
physics program at a $\phi$ factory.

\SECTION{\label{ack} Acknowledgements}
 
The author thanks Tullio Bressani, Giovanni Carboni, Paolo Franzini,
Luciano Paoluzi and Rinaldo Santonico for many useful discussions and comments.
The author particularly acknowledges the encouragement and 
most valuable advices of Giorgio Matthiae. 
 
\pagebreak

\pagebreak
\newpage

\begin{figure}[p]
\mbox{\epsfysize20.0cm\epsffile{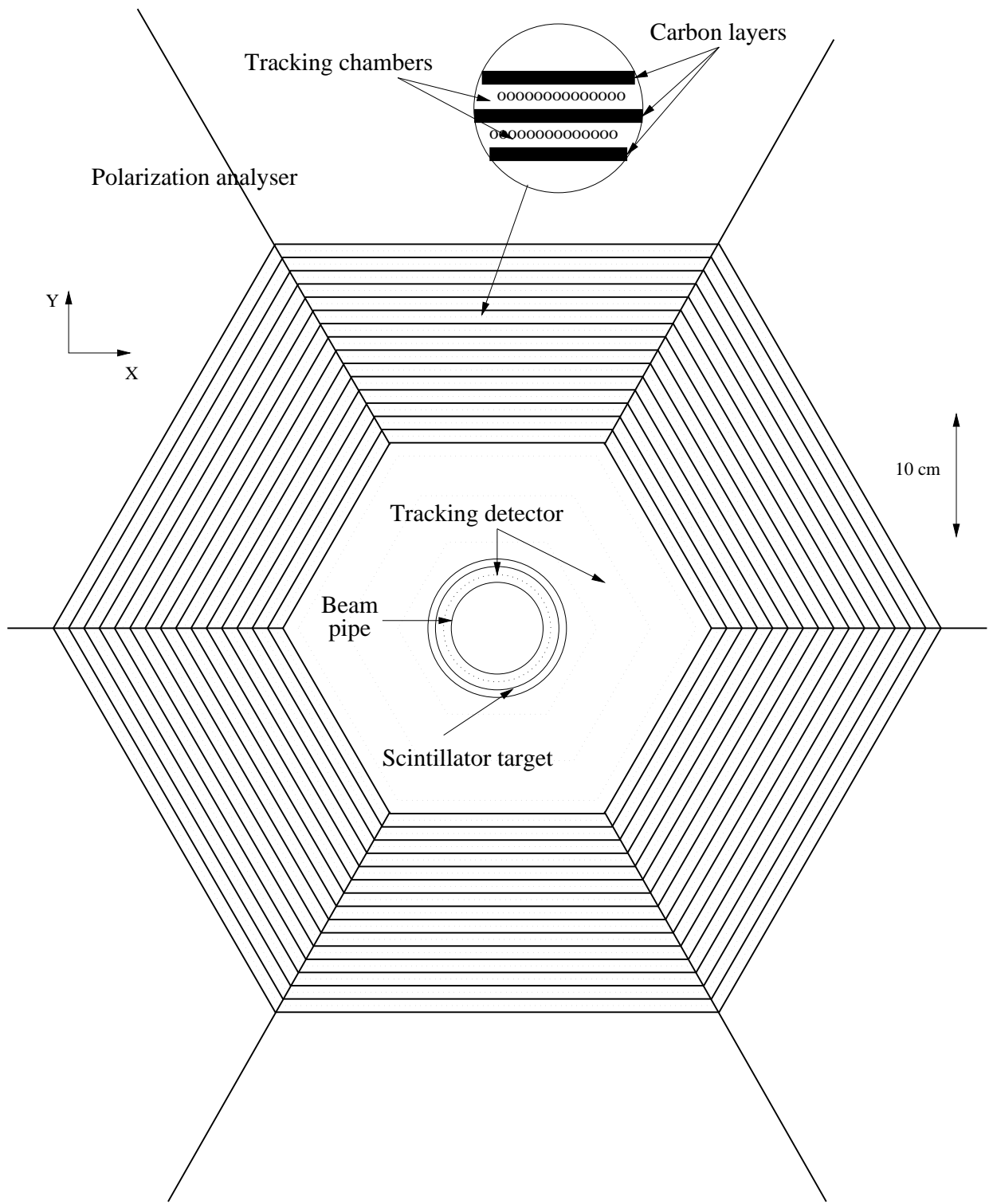}}
\caption[Etika detector]{Schematic picture of the ETIKA experiment} 
\label{fig:etikd}
\end{figure}

\end{document}